\documentclass[prb,aps,tightenlines,twocolumn,reprint,
showpacs,floatfix]{revtex4}
\usepackage{graphicx}
\usepackage{amssymb}
\usepackage{amsmath}
\usepackage{bm}
\usepackage{colordvi}
\usepackage{mathrsfs}
\usepackage{cases}
\def\up{\uparrow}
\def\down{\downarrow}

\begin{document}
\title{Majorana fermions in $T$-shaped semiconductor nanostructures}
\author{Y. Zhou}
\email{zya2538@mail.ustc.edu.cn}
\author{M. W. Wu}
\email{mwwu@ustc.edu.cn}
\affiliation{Hefei National Laboratory for Physical Sciences at
  Microscale and Department of Physics, University of Science and
  Technology of China, Hefei, Anhui, 230026, China}

\date{\today}
\begin{abstract}
We investigate the Majorana fermions in a $T$-shaped semiconductor
nanostructure with the Rashba spin-orbit coupling and a magnetic 
field in the proximity of an s-wave superconductor. It is found that the
properties of the low-energy modes (including the Majorana and near-zero-energy
modes) at the ends of this system are similar to those in the Majorana
nanowire. However, very distinct from the  
nanowire, one Majorana mode emerges at the intersection of the $T$-shaped
structure when the number of the low-energy modes at each end $N$ is odd,
whereas neither Majorana nor near-zero-energy mode appears at the intersection
for even $N$. 
We also discover that the intersection Majorana mode plays an important role in
the transport through the above $T$-shaped nanostructure with each end connected
with a normal lead. 
Due to the presence of the intersection mode, the deviation of the zero-bias
conductance from the ideal value in the long-arm limit $Ne^2/h$ is more
pronounced in the regime of odd $N$ compared to the one of even $N$. 
Furthermore, when the magnetic field increases from the 
regime of odd $N$ to the one of even $N+1$, the deviation 
from the ideal value tends to decrease.
This behavior is also very distinct from that in a nanowire, where the 
deviation always tends to increase with the increase of magnetic field.
\end{abstract}



\maketitle

\section{Introduction}
Since Majorana fermions, particles which are their own antiparticles, are
proposed theoretically in topological superconductors,\cite{Moore_91,Ivanov_01, 
Kitaev,Fu,Sarma_vortex,Sarma_single,Sarma_ring,Oreg_10,Stanescu_rev,
Flensberg_rev,Alicea_rev} 
the search of Majorana fermions has attracted much attention in the condensed
matter community.
Apart from the importance for fundamental physics, Majorana modes in topological
superconductors are of great use for quantum computation due to their non-Abelian
exchange statistics.\cite{Ivanov_01,Read_00,Sarma_rev,Alicea_11}

A promising proposal for engineering Majorana quasiparticles is based on
a semiconductor nanowire with both spin-orbit coupling (SOC) and magnetic
field in the proximity of an s-wave
superconductor.\cite{Sarma_single,Sarma_ring,Oreg_10,Stanescu_rev} 
In a long one-dimensional nanowire, in which only the lowest subband is
involved, Majorana modes emerge as one pair of zero-energy states located at 
the two ends of the nanowire in the parameter regime satisfying the 
condition $|V_{\rm Z}|>\sqrt{\Delta^2+\mu^2}$.\cite{Sarma_single} 
Here $V_{\rm Z}$, $\Delta$ and $\mu$ represent the Zeeman splitting induced by
the magnetic field, proximity-induced superconducting gap and chemical
potential, respectively.  A characteristic feature of Majorana 
states is the conductance peak at zero bias.\cite{Law_resonant,Flensberg_NS,
Prada_NS,Ioselevich_discrete,Sarma_ZBP,Beenakker_NS_Q,Lim_trans_NSN,Wang_NSN,
Mourik_exp,Deng_exp,Das_exp,Churchill_exp,Roy_ZBP_true} The value of
this zero-bias peak is predicted to be quantized, i.e., $2e^2/h$ for 
a normal-superconductor surface\cite{Law_resonant,Flensberg_NS,Prada_NS,
Ioselevich_discrete,Sarma_ZBP,Beenakker_NS_Q} and $e^2/h$
for a normal-superconductor-normal (NSN) structure, as the NSN
structure consists of two normal-superconductor
surfaces.\cite{Lim_trans_NSN,Wang_NSN} 
When the length of wire is comparable or shorter than the coherence length of
Majorana modes, the interaction between the two end Majorana modes becomes
important and leads to an energy splitting of these states.\cite{Sarma_ZBP,Sarma_smoking} 
This effect can reduce the value of the zero-bias peak and can even make this peak
split into two peaks at finite bias when the splitting is large
enough.\cite{Ioselevich_discrete,Sarma_ZBP} 

There are also a few works on Majorana fermions in multi-subband
nanowires.\cite{Potter_multi,Lutchyn_multi,Sarma_multi,Lim_mixing,
Lim_trans_NSN,Gibertini_multi,Sarma_multi2,Beenakker_NS_Q} 
This system supports multiple low-energy modes at each ends.
The number of these modes $N$ is determined by the $Z$ topological
invariant, which comes from the approximate chiral
symmetry.\cite{Beenakker_NS_Q,Sarma_multi2,Stanescu_rev} 
In the weak superconducting-pairing limit, $N$ is approximately equal to the
number of the subbands in which only the states with 
one kind of spin are occupied.\cite{Sarma_multi,Sarma_multi2,Stanescu_rev} 
Without considering the SOC between different transverse subbands, the chiral
symmetry is exact, and hence all the low-energy modes at the ends are the
Majorana modes in the long-wire limit. 
Nevertheless, taking account of the inter-subband SOC, the chiral symmetry is
weakly broken, and most of the low-energy modes are split off and become the
near-zero-energy modes.
The number of the left Majorana modes is determined by the $Z_2$ topological
invariant, which corresponds to the parity of
$N$.\cite{Sarma_single,Sarma_multi2,Stanescu_rev} 
In the nontrivial (trivial) topological regime with odd (even) $N$, there is one
(no) Majorana mode at each end. 
The presence of the Majorana mode for odd $N$ is also according to the
restriction from the particle-hole symmetry of the Bogoliubov-de Gennes (BdG) 
Hamiltonian. When the splittings of these low-energy modes are still negligible
compared with the energy broadening from the leads, the near-zero-energy modes
behave the same as the Majorana modes. 
Thus, the conductance still shows a peak at zero bias with
peak value being $Ne^2/h$ in an NSN structure. 
However, when the splitting induced by the inter-subband SOC becomes important,
the behavior of the conductance becomes complex due to the interference between
different low-energy modes and the zero-bias conductance can vary between 0 and
$Ne^2/h$.\cite{Lim_trans_NSN}

So far, all works in this field focus on the two-terminal Majorana nanowire.
There is no report on the three-terminal Majorana
nanostructure, eg., the $T$-shaped semiconductor nanostructure with
the Rashba SOC and magnetic field in the proximity of superconductor. 
In fact, the three-terminal $T$-shaped structure built with normal conductor has
been extensively investigated and shows very distinct electric and transport
properties. In particular, a localized state appears at the intersection of the
three-terminal structure and induces the Fano line shape in the bias dependence
of the conductance.\cite{Schult_T,Openov_T,Lin_T,Goldoni_T,Baranger_T,
Chen_T,Bohn_T,Xu_T_GNR} 
Thus, it is expected that there are also some interesting features in the 
$T$-shaped Majorana nanostructure. Furthermore, the unique  property of this
system can be shown in a simple view. We take the case where both the 
main-arm and side-arm are one-dimensional chains as an example.
Due to similarity of the Hamiltonian around the ends of the $T$-shaped Majorana
nanostructure and nanowire, the Majorana modes should appear at all three ends
of the $T$-shaped nanostructure in the nontrivial topological regime. 
Thus, the total number of zero-energy end states is three. 
However, it is known that Majorana modes must emerge in pairs, since one
Majorana fermion only contains half the degrees of freedom of a normal 
fermion.\cite{Flensberg_rev,Alicea_rev} 
Therefore, an additional unknown Majorana mode must exist.
The main purpose of this work is to identify this additional Majorana mode 
and reveal its influence on the transport through the $T$-shaped Majorana
nanostructure.

In this paper, we investigate the low-energy spectrum and transport 
properties of a $T$-shaped Majorana nanostructure. 
We discover that, distinct from the behavior in nanowires, where all
Majorana modes appear at the two ends, a Majorana mode shows up at the
intersection of the three-terminal $T$-shaped structure in the case with odd
low-energy modes (including the zero-energy and near-zero-energy modes) at the
three ends. However, neither Majorana  nor near-zero-energy
mode appears 
at the intersection in the case with even low-energy modes at the three ends.
We further show that, the presence of the intersection Majorana mode 
enhances the deviation of the zero-bias conductance from its ideal 
value in the long-arm limit. 
Also considering that the regime with the intersection Majorana mode can
appear at lower magnetic field compared to that without it, the deviation from 
the ideal value tends to decrease with the increase of magnetic field in this
case. This behavior is also distinct from the transport property through a
Majorana nanowire, where the deviation always shows increasing trend
with increasing magnetic field.\cite{Lim_trans_NSN}

The paper is organized as follows. In Sec.~II, we set up the tight-binding
Hamiltonian of a $T$-shaped Majorana nanostructure and calculate the low-energy 
spectrum and identify the Majorana states.
In Sec.~III, we derive the formula of current between different ends of the
$T$-shaped structure and present the numerical results of electric conductance.
Finally, we conclude in Sec.~IV.

\section{Hamiltonian and Energy Spectrum}
\subsection{Hamiltonian}
\begin{figure}[tbp]
  \begin{center}
    \includegraphics[width=8.5cm]{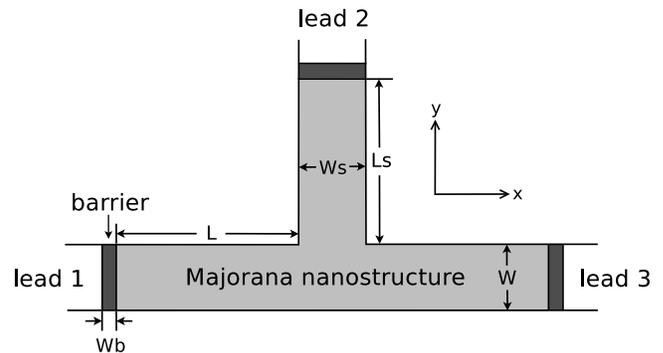}
  \end{center}
  \caption{Schematic view of a $T$-shaped Majorana
    nanostructure with each end connected with a normal lead.
  }
  \label{fig_structure} 
\end{figure}

We consider a $T$-shaped semiconductor nanostructure with the Rashba SOC and
proximity-induced superconducting pairing in the presence of a
magnetic field perpendicular to the plane of this structure,
as sketched in Fig.~\ref{fig_structure}.
Note that the leads plotted in this figure are excluded in this section but
included in calculating the transport properties in the next section. 
The tight-binding Hamiltonian of this structure can be written
as\cite{Sarma_multi,Potter_multi,Gibertini_multi} 
\begin{eqnarray}
  H_{\rm eff}&=& {H}_0+H_{\rm SC},\\
  {H}_0&=& \sum_{i\sigma}( {\sigma} 
  V_{\rm Z}+V_{i}- \mu) c_{i\sigma}^\dagger c_{i\sigma} 
  - \sum_{\langle i,j \rangle \sigma} t c_{i\sigma}^\dagger c_{j\sigma} 
  \nonumber \\
  && {}+{i E_R} \sum_{\langle i,j \rangle\sigma\sigma'} 
  (v^y_{ij}\sigma^x_{\sigma \sigma'}-v^x_{ij}\sigma^y_{\sigma \sigma'}) 
  c_{i \sigma}^\dagger c_{j \sigma'}, 
  \label{Hami_0}\\
  H_{\rm SC}&=& \sum_{i} \Delta c_{i+}^\dagger
  c_{i-}^\dagger + {\rm H. c.}.
\end{eqnarray}
Here $t$ represents the hoping energy; $V_{i}=4t$ denotes the on-site energy;
$\langle i,j \rangle$ represents a pair of the nearest neighbors;
$\sigma^l$ for $l=x,y$ are the Pauli matrices;
$v_{ij}^l={\bf e}_l \cdot {\bf d}_{ij}$ with 
${\bf d}_{ij}=({\bf r}_i-{\bf r}_j)/|{\bf r}_i-{\bf r}_j|$;
$E_R$ represents the Rashba SOC constant.

It is convenient to rewrite the above Hamiltonian into the following
form\cite{BdG_original} 
\begin{equation}
  H_{\rm eff}=\frac{1}{2} \sum_{ij} \bar{\Phi}_i^\dagger 
  {H}_{{\rm BdG}}(i,j) \bar{\Phi}_j,
\end{equation}
where $\bar{\Phi}_i^\dagger= \left( c^\dag_{i\up}, c^\dag_{i\down},
c_{i\down}, -c_{i\up} \right)$ denotes the Nambu spinor and
\begin{equation}
  {H}_{\rm BdG}(i,j)=\begin{pmatrix} {H}_{0}(i,j) && \Delta \delta_{ij} \\
  \Delta^* \delta_{ij} && -{\sigma}_y {H}_{0}^\ast(i,j) {\sigma}_y 
  \end{pmatrix}
\end{equation}
represents the BdG Hamiltonian.\cite{BdG_original}
By diagonalizing the matrix form of BdG Hamiltonian (labelled by
$\hat{H}_{\rm BdG}$), one obtains the energy spectrum and eigenstates
of this system.

\begin{figure}[tbp]
  \begin{center}
    \includegraphics[width=7.2cm]{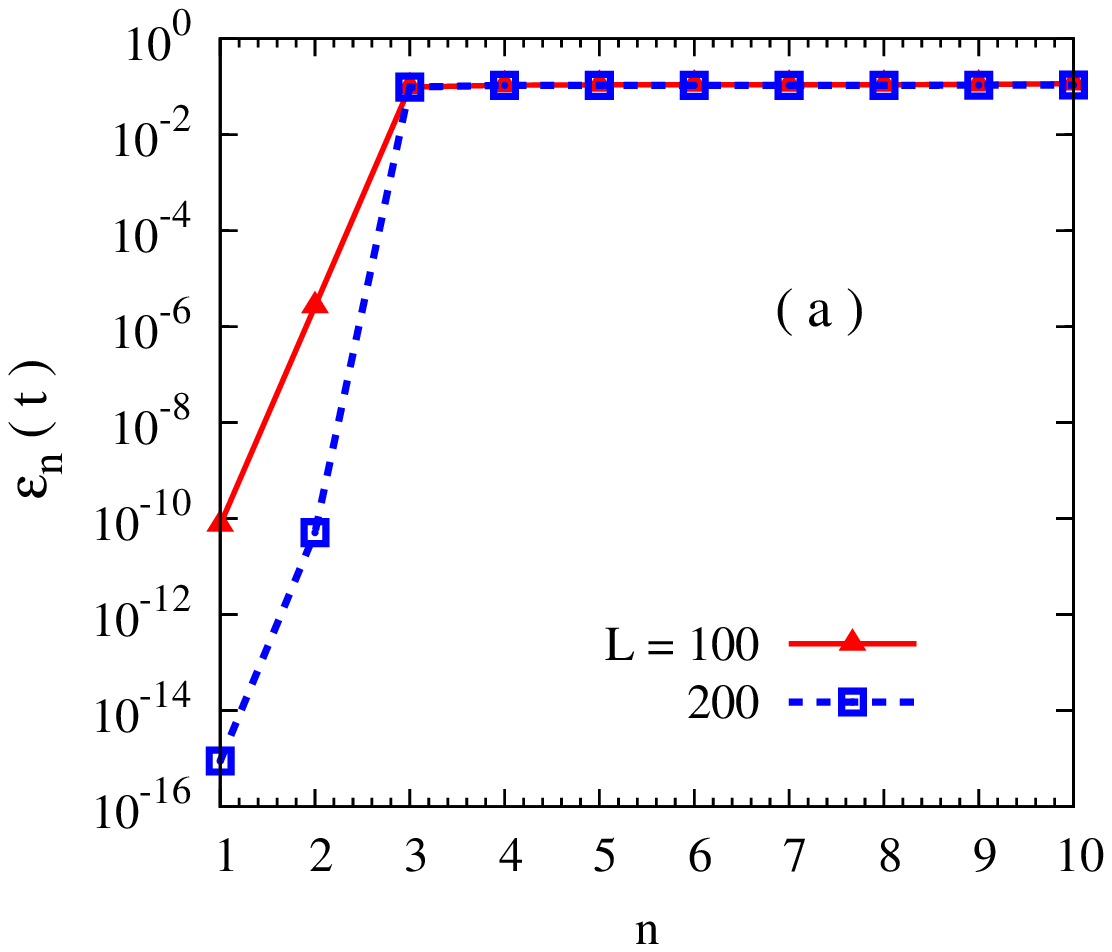}
    \includegraphics[width=8.cm]{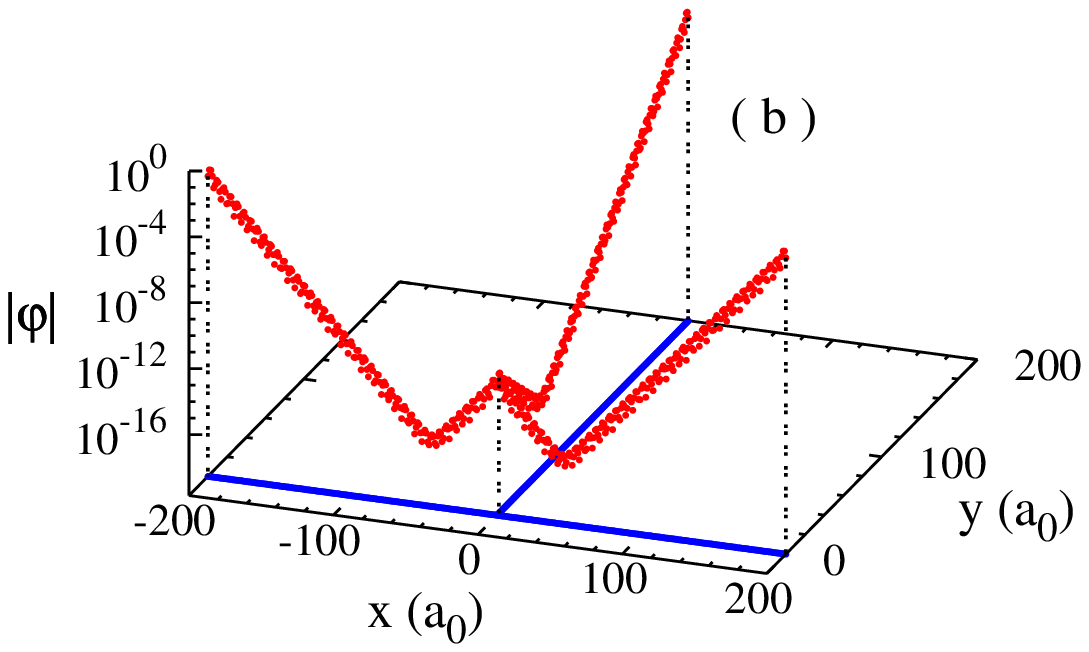}
    \includegraphics[width=8.cm]{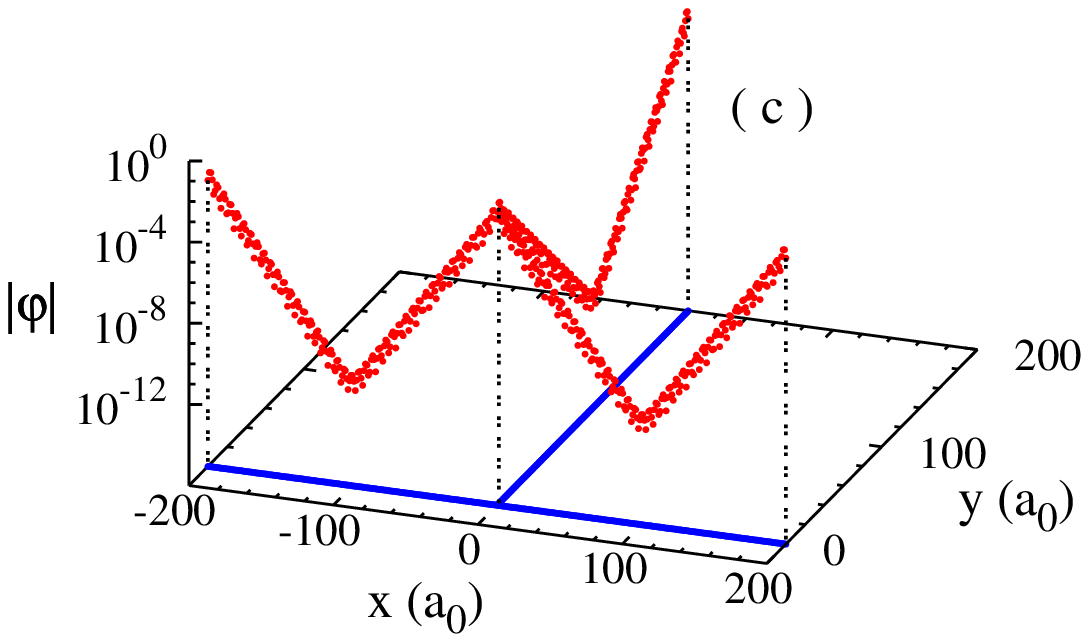}
  \end{center}
  \caption{(Color online) Isolated $T$-shaped Majorana nanostructures
    with $W=W_s=1$, $\mu=E_0$ and $V_{\rm Z}=0.4 t$. (a) Low energy spectra for
    $L=L_s=100$ and $200$. $n$ labels the eigenvalues of $H_{\rm eff}$ staring  
    with zero energy. The curves are only plotted as a guide for the eye.
    (b) (c) Magnitude of the wave functions of the lowest two eigenstates for
    $L=L_s=200$. 
  }
  \label{fig_band1} 
\end{figure}

\subsection{Low-energy spectrum and Majorana states} 
In this subsection, we present the numerical results of the low-energy spectrum
and eigenstates. We choose $\Delta=E_R=0.2 t$ unless otherwise specified. 
We first discuss the simplest $T$-shaped Majorana nanostructure with
$W=W_s=1$. Evidently, there is only one subband in this case. 
As mentioned in the introduction, due to the similarity of this system and the
one-dimensional nanowire, the Majorana modes are expected to appear at all three 
ends of the $T$-shaped nanostructure in the nontrivial topological regime, 
i.e., $\sqrt{\Delta^2+(\mu-E_0)^2}<|V_{\rm Z}|<
\sqrt{\Delta^2+(\mu-E_0-4t)^2}$, where $E_0$ represents the energy of bulk
states at $k_{\parallel}=0$ with $k_{\parallel}$ being the momentum along the
free propagating direction.\cite{upper_limit} 
Nevertheless, there must be addition Majorana mode(s) to ensure that the total
number of Majorana modes are even.
To confirm the above claim, we plot the low-energy spectrum of this structure
with $L=L_s=100$ and $200$ for $\mu=E_0$ and $V_{\rm Z}=0.4 t$, which
belongs to the nontrivial topological regime, in
Fig.~\ref{fig_band1}(a). 
Here only the ones in the positive eigenenergy regime are shown due to the
particle-hole symmetry of the BdG Hamiltonian. 
It is seen that there are two eigenstates
with extremely small energy and their energies decrease almost
exponentially with the increase of the arm length. This indicates that
both states are strict zero-energy states in the long-arm limit. 
Considering one zero-energy fermonic state corresponds to one pair of Majorana
fermions,\cite{Flensberg_rev,Alicea_rev} 
one finds that there are four (two pairs of) Majorana modes in total.
This means that there is indeed an additional unknown Majorana mode.
To identify this additional Majorana mode, we further plot the magnitude of the
wave functions of the two zero-energy eigenstates for $L=L_s=200$ in
Figs.~\ref{fig_band1}(b) and (c), respectively. The results show that the
additional Majorana mode is located exactly at the intersection. Note that both
zero-energy eigenstates include the contribution from all four Majorana modes,
since the Majorana states are degenerate and the coupling between them is still
finite in the finite-sized system. 
The emergence of the intersection Majorana mode can
be understood through the analytic formula of the Majorana modes in
this system (see Appendix~\ref{MF_spe}).

\begin{figure}[tbp]
  \begin{center}
    \includegraphics[width=8.5cm]{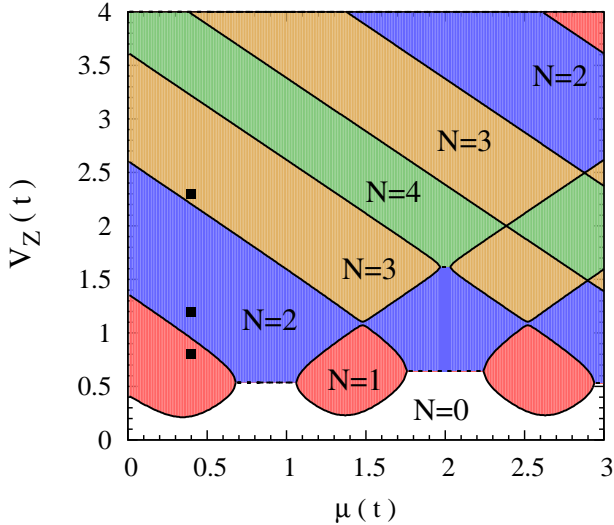}
  \end{center}
  \caption{(Color online)  Phase diagram in an isolated $T$-shaped Majorana
    nanostructure with $W=W_s=4$ as function of the Zeeman splitting 
    $V_{\rm Z}$ and the chemical potential $\mu$. $N$ represents the number of
    the low-energy modes at each end. 
    The black squares indicate the chemical potential and Zeeman splitting
    used in Fig.~\ref{fig_multi}.
  }
  \label{fig_phase} 
\end{figure}

\begin{figure}[thbp]
  \begin{center}
    \includegraphics[width=4.cm,height=3.5cm]{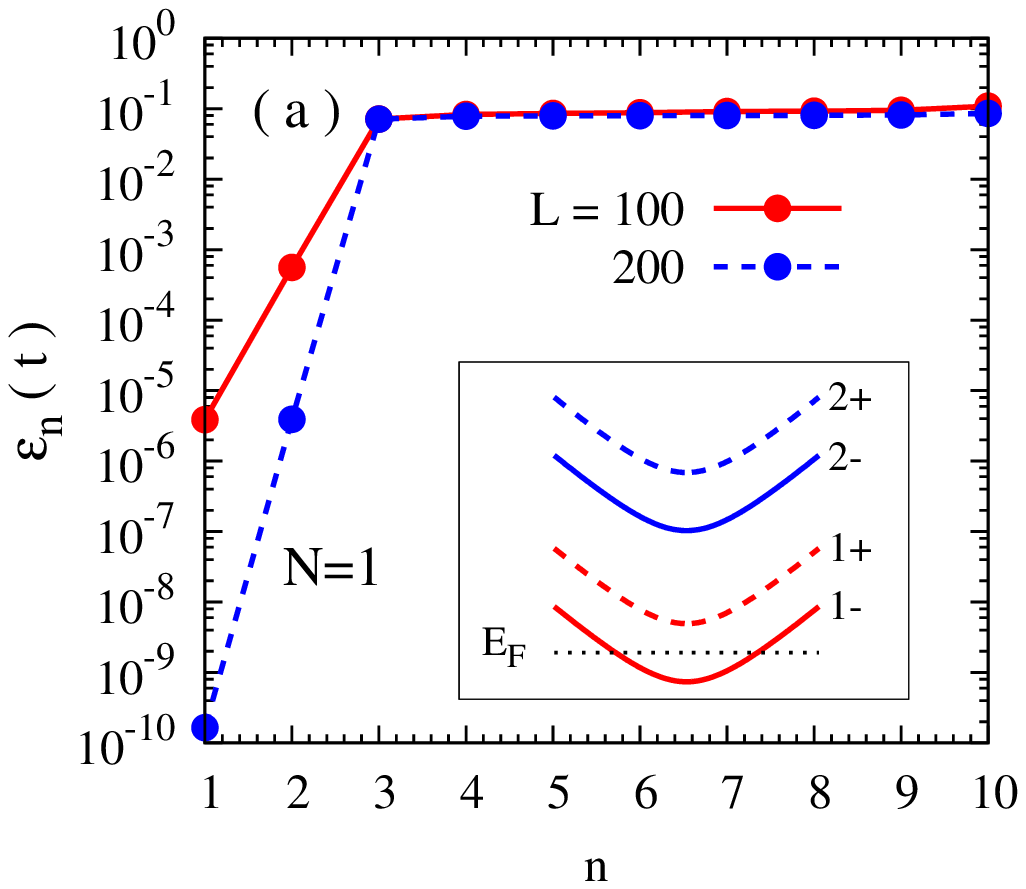}
    \includegraphics[width=4.5cm,height=3.5cm]{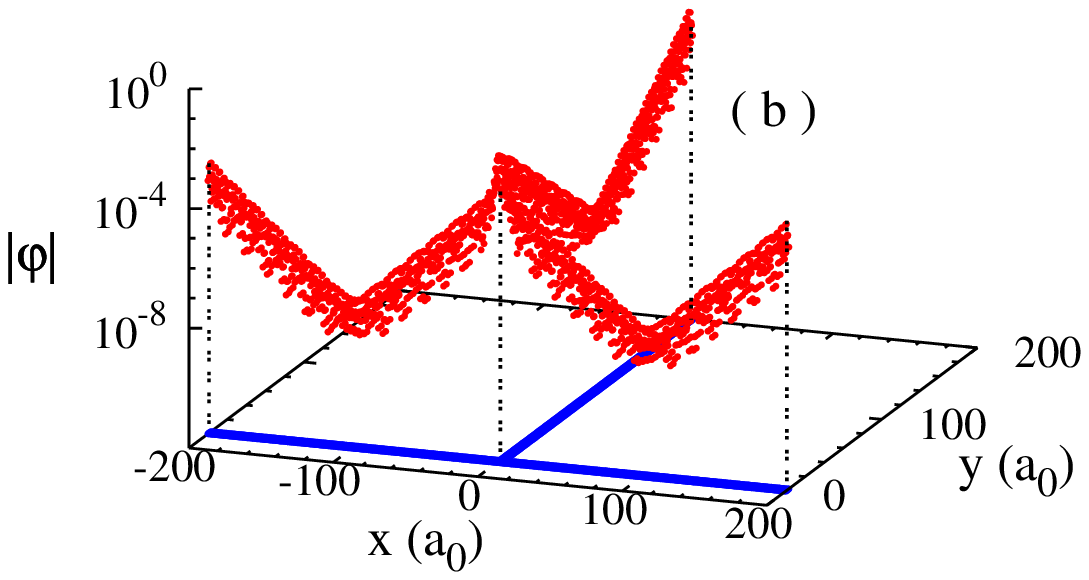}
    \includegraphics[width=4.cm,height=3.5cm]{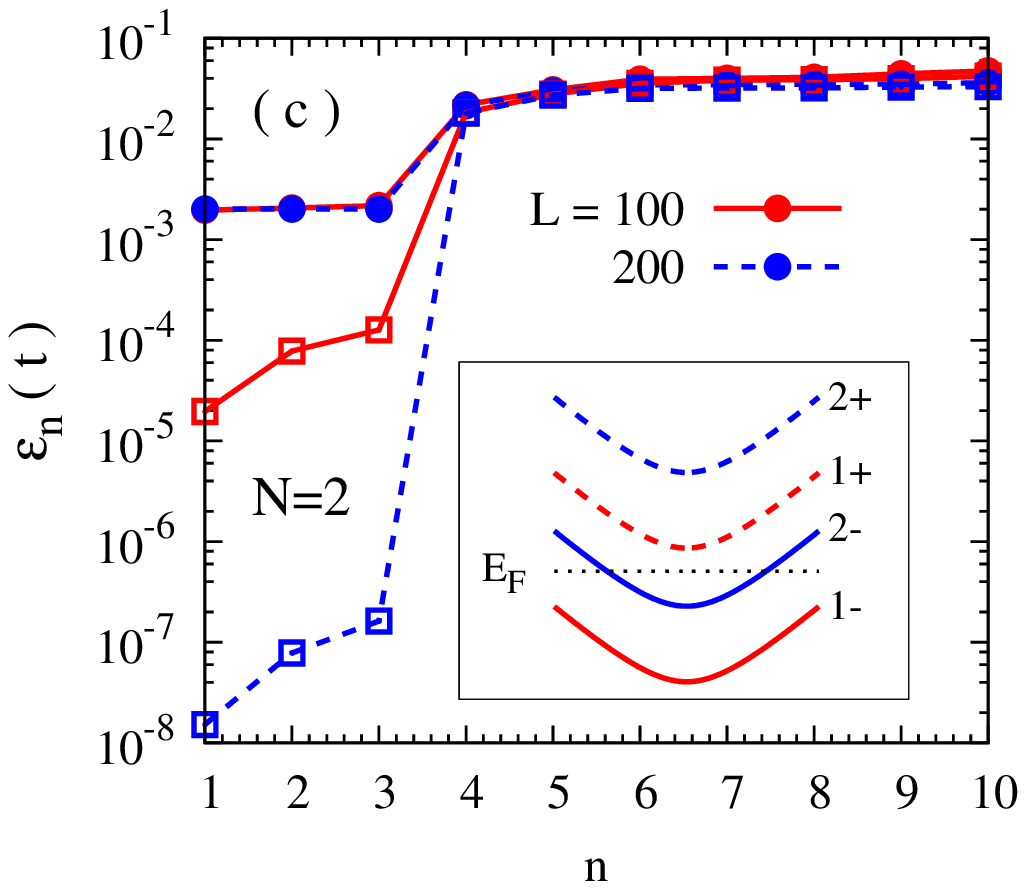}
    \includegraphics[width=4.5cm,height=3.5cm]{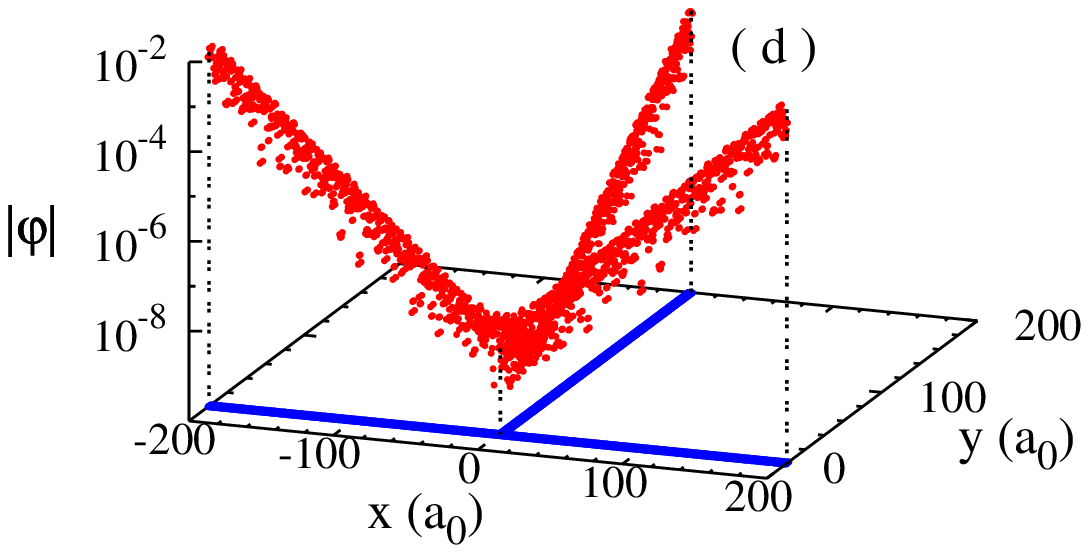}
    \includegraphics[width=4.cm,height=3.5cm]{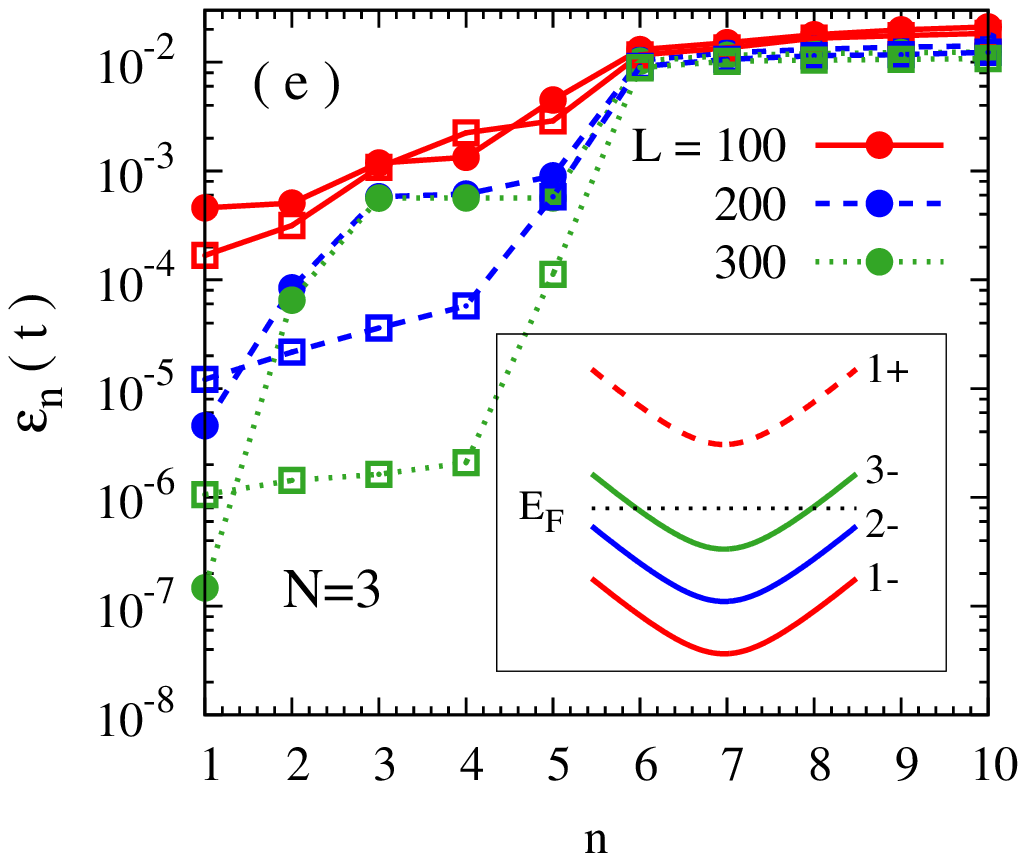}
    \includegraphics[width=4.5cm,height=3.5cm]{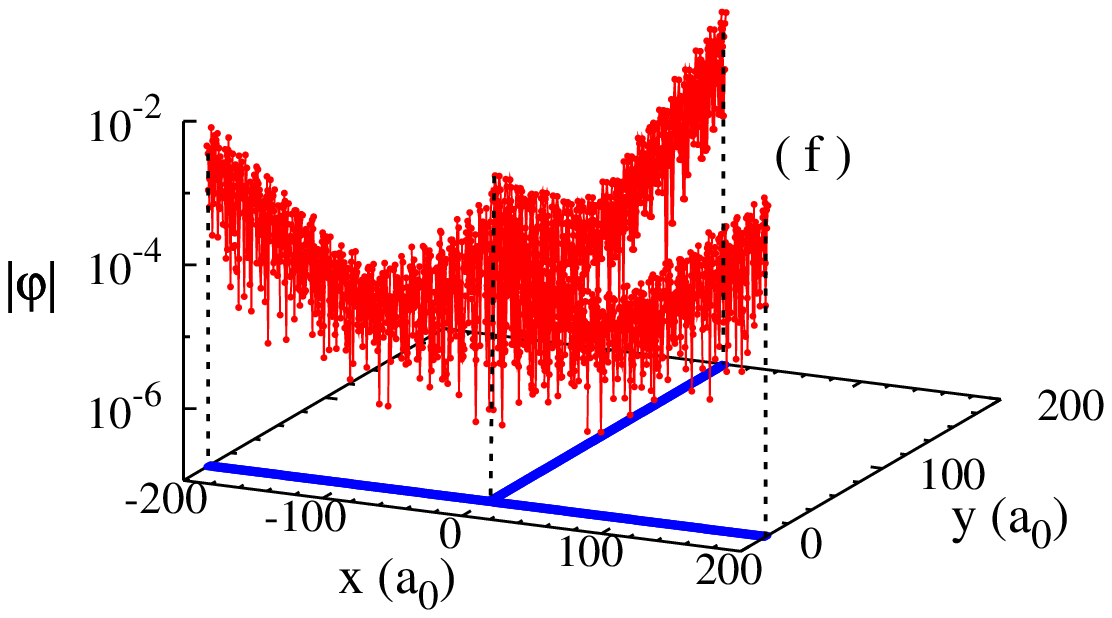}
  \end{center}
  \caption{\Black{(Color online) Isolated $T$-shaped Majorana nanostructures
    with $W=W_s=4$. $\mu=0.4 t$; $V_{\rm Z}=0.8 t$ (a) (b), $1.2t$ (c) 
    (d), $2.3 t$ (e) (f).
    (a) (c) (e) Low-energy spectra for different arm lengths $L=L_s$.
    The dots and squares represent the results with and without the
    inter-subband SOC, respectively. 
    The curves are only plotted as a guide for the eye. 
    In the insets, we also schematically plot the occupation of the four lowest
    spin bands.
    (b) (d) (f) Magnitude of the wave functions of the low-energy eigenstates
    with $n=2$ for $L=L_s=200$.}
  }
  \label{fig_multi} 
\end{figure}

We then turn to the multi-subband $T$-shaped Majorana nanostructure. 
The number of the low-energy modes $N$ at each end of this structure
can be obtained following the similar approach in the multi-subband
nanowire.\cite{Sarma_single,Beenakker_NS_Q,Sarma_multi2,Stanescu_rev} 
It is determined by the $Z$ topological invariant from the approximate chiral
symmetry.\cite{Beenakker_NS_Q,Sarma_multi2,Stanescu_rev} 
In most of the parameter regime investigated here, the superconducting pairing
is much weaker than the chemical potential and Zeeman splitting. In these
cases, $N$ is approximately equal to the number of the subbands in which only
the states with one kind of spin are occupied. 
However, as the inter-subband SOC weakly breaks the chiral symmetry,
most of these low-energy modes are only the near-zero-energy modes instead of
the Majorana modes.\cite{Sarma_multi2,Stanescu_rev} 
The number of the Majorana modes is determined by the $Z_2$ 
topological invariant, which corresponds to the parity of
$N$.\cite{Sarma_single,Sarma_multi2,Stanescu_rev} 
One (no) Majorana mode appears at each end 
in the nontrivial (trivial) topological regime with odd (even) $N$.

In Fig.~\ref{fig_phase}, we plot the phase diagram of the $T$-shaped
Majorana nanostructure with $W=W_s=4$.
Here the regions with the same color share the same $N$. 
The solid curves represent the transition points of the $Z_2$ topological
invariant, which are obtained from the gap closing 
condition of the bulk energy spectrum at $k_{\parallel}=0$ or
$\pi/a$.\cite{Sarma_single,Gibertini_multi} 
It is observed that these solid regime boundaries show some anti-crossings, 
eg., between the regimes with $N=0$ and $2$.
This effect comes from the anti-crossings between the energy spectra of bulk
states in different subbands, which is induced by the inter-subband SOC.
This kind of anti-crossings are also observed in the phase diagram in the
quasi-1D nanowire with magnetic field 
perpendicular to the nanowire plane.\cite{Stanescu_rev,Potter_multi} 
Around these anti-crossings, the inter-subband SOC cannot be treated
perturbatively, thus the $Z$ topological invariant and $N$ cannot be well
defined. This indicates that there are no strict 
boundaries between the relevant regimes. 
Here we only plot dashed curves at the positions where the bulk gap at
$k_{\parallel}=0$ or $\pi/a$ reaches a finite minimum as rough boundaries to
separate these regimes inside these anti-crossings. 

Now we examine the existence of the Majorana or near-zero-energy modes in
different regimes for $W=W_s=4$. 
We choose $\mu=0.4 t$ and $V_{\rm Z}=0.8 t$, $1.2 t$ and $2.3 t$, which
belong to the regimes with $N=1$, 2 and 3, respectively, 
as indicated by the squares in Fig.~\ref{fig_phase}. The energy spectra in
these three cases are plotted in Figs.~\ref{fig_multi}(a), (c) and (e),
respectively. 
In the insets of these figures, we also schematically plot the occupation of the
four lowest spin bands, in which the $m\sigma$ band represents the spin-majority
($\sigma=-$) or -minority ($\sigma=+$) band of the $m$-th transverse subband.
We first focus on the case with $N=1$, where only the spin-majority
band of the lowest subband ($1-$) is occupied. 
It is seen that both the energy spectrum [Fig.~\ref{fig_multi}(a)] and the wave
functions of the low-energy states [the one with $n=2$ for $L=L_s=200$ is shown
in Fig.~\ref{fig_multi}(b)] in this case are similar to those in the $T$-shaped
nanostructure with $W=W_s=1$ discussed above: there are four (two pairs of)
Majorana modes in total and one appears at the intersection. 

Then we turn to the case with $N=2$ [curves with dots in
Fig.~\ref{fig_multi}(c)], where
the spin-majority bands of the lowest two subbands ($1-$ and $2-$) are
occupied. It is shown that there are three low-energy eigenstates. In the 
long-arm limit, their energies become very close to 
each other and all saturate to the order of $10^{-3} t$. This indicates that
they are only the near-zero-energy states but not the Majorana states.
It is also seen that, after removing the inter-subband SOC (curves with
squares),\cite{interband} the energies of all the low-energy states 
decrease with the increase of arm length and hence recover the behavior of the
Majorana modes. This further justifies that the small splitting of these
near-zero-energy states is due to the inter-subband SOC, in consistence 
with the above discussions based on the topological invariant.
The absence of the Majorana modes also agrees with the fact that
the regime with even $N$ belongs to the trivial topological phase.
Moreover, we also plot the wave function of the eigenstate with $n=2$ for
$L=L_s=200$ in Fig.~\ref{fig_multi}(d). The wave 
functions of the other two low-energy eigenstates are similar to this one. 
One finds that all low-energy eigenstates are constructed by the
near-zero-energy modes at the ends and hence the intersection near-zero-energy 
mode does not appear.

The behavior for $N=3$ [curves with dots in Fig.~\ref{fig_multi}(e)], in which
the $1-$, $2-$ and $3-$ bands are occupied, is more complex. It is found that 
there are five low-energy eigenstates. The lowest two tend to be
zero energy with increasing length and the other three saturate to the order of 
$10^{-3} t$, indicating two zero-energy eigenstates and three near-zero-energy
ones in the long-arm limit. The presence of the Majorana fermions is consistent
with the fact that the regime with odd $N$ belongs to the nontrivial topological
phase. The finite splitting of the near-zero-energy states also comes from the
inter-subband SOC, as indicated by the comparison of the energy spectra with
(curves with dots) and without the inter-subband SOC (curves with squares). 
We further examine the wave functions of the zero-energy eigenstates and plot
the one with $n=2$ in Fig.~\ref{fig_multi}(f). It is shown that 
there is also a Majorana mode at the intersection, just similar to the
case with $N=1$. In addition, we verify that there is no near-zero-energy
mode at the intersection in this case.
Based on the above discussions, one can conclude that one intersection Majorana
mode appears in the case with odd $N$, while there is neither Majorana nor
near-zero-energy mode at the intersection in the case with even $N$.

\section{Electric Conductance}
\subsection{Formalism}
In this section, we discuss the transport properties through the $T$-shaped
Majorana nanostructure and, more importantly, identify the role of the
intersection Majorana modes in it. 
Here we connect each end of this structure with a normal
lead, as indicated in Fig.~\ref{fig_structure}. 
We also add barriers between the leads and the $T$-shaped structure to reduce the
broadening of energy level and hence avoid the states above the bulk gap
contributing to the low-energy transport. 
The Hamiltonian of the leads (including the barriers) is
similar to $H_0$ in the $T$-shaped structure and can be written as
\begin{eqnarray}
  {H}_\eta&=& 
  \sum_{i \sigma}( {\sigma} V_{\rm Z}+V_{i}- \mu) 
  d_{\eta i\sigma}^\dagger d_{\eta i\sigma} 
  - \sum_{\langle i,j \rangle \sigma} t 
  d_{\eta i\sigma}^\dagger d_{\eta j\sigma} \nonumber \\
  &&\hspace{-0.3cm} {} + {i E_R}
  \sum_{\langle i,j \rangle \atop \sigma\sigma'}
  (v^y_{i j}\sigma^x_{\sigma \sigma'}
  -v^x_{i j}\sigma^y_{\sigma \sigma'}) 
  d_{\eta i \sigma}^\dagger d_{\eta j \sigma'},
\end{eqnarray}
where $\eta=1,2,3$ represents the left, central and right leads (see
Fig.~\ref{fig_structure}), 
the on-site energy $V_{i}$ is chosen to be  $4t+V_b$ 
with $V_b$ being the barrier height
 in the barrier region and $4t$ otherwise. The
hoping between the leads and the $T$-shaped structure is described by 
\begin{eqnarray}
  {H}_T&=& \sum_{\eta \langle i,j \rangle \atop \sigma\sigma'}
  T^{\eta}_{i\sigma,j\sigma'} d_{\eta i \sigma}^\dagger c_{j \sigma'}, \\
  T^{\eta}_{i\sigma,j\sigma'}&=& -t \delta_{\sigma \sigma'}
  + {i E_R}(v^y_{i j}\sigma^x_{\sigma \sigma'}
  -v^x_{i j}\sigma^y_{\sigma \sigma'}).
\end{eqnarray}


The electric current flowing away from the lead $\eta$ can be
written as 
\begin{equation}
   I_\eta(t)=e\partial_t\langle \mathcal{N}_\eta(t)\rangle
    =\frac{ie}{\hbar}\langle [H(t),\mathcal{N}_\eta(t)]\rangle \qquad
\end{equation}
with $\mathcal{N}_\eta(t)=\sum\limits_{i}d_{\eta i\sigma}^\dag(t) 
d_{\eta i\sigma}(t)$.
Using the Green's functions in the Nambu spinor basis (see
Appendix~\ref{Green_Func}) and following the similar way deriving the current
through normal mesoscopic nanostructures,\cite{Haug_1998} one obtains
\begin{equation}
  I_\eta = \frac{e}{h} \int_{-\infty}^{\infty} d\varepsilon \sum_{\eta'\beta}
  P_{\eta\eta'}^{e\beta}(\varepsilon) [f_{\eta e}(\varepsilon)
    -f_{\eta'\beta}(\varepsilon)], 
  \label{current}
\end{equation}
in which
\begin{equation}
  P^{\alpha\beta}_{\eta\eta'}(\varepsilon)={\rm Tr} \left\{
    \hat{G}^{r}(\varepsilon) \hat{\Gamma}^{\beta}_{\eta'}(\varepsilon) 
    \hat{G}^{a}(\varepsilon)
    \hat{\Gamma}^{\alpha}_{\eta}(\varepsilon) \right\}. 
  \label{P_def}
\end{equation}
Here $\hat{G}^{r,a}(\varepsilon)$ are the retarded and advanced Green's functions 
in the $T$-shaped nanostructure connected with leads [see
Eq.~(\ref{Dyson_c})]; $\hat{\Gamma}^{\alpha}_{\eta}(\varepsilon)$ is the
self-energy from the electric ($\alpha=e$) or hole part ($\alpha=h$) in the lead
$\eta$ [see Eq.~(\ref{Sigma_def})].
It is noted that this formula of current is just equivalent to the one obtained
from the transfer matrix approach.\cite{Lambert_trans_super,Lim_trans_NSN}

At zero temperature, Eq.~(\ref{current}) can be reduced into
\begin{equation}
  I_\eta =\frac{e}{h} \sum_{\eta'\beta}
  \int_{\chi_\beta\mu_{\eta'}}^{\mu_\eta} d\varepsilon \;
  P_{\eta\eta'}^{e\beta}(\varepsilon),
  \label{current_2}
\end{equation}
where $\chi_\beta=1$ ($-1$) for $\beta=e$ ($h$) and $\mu_\eta$ is the chemical 
potential in lead $\eta$. The differences between chemical potentials in
different leads are determined by the bias $eV_1=\mu_1-\mu_3$ and
$eV_2=\mu_2-\mu_3$. 
In this investigation, we focus on two quantities: (i) the differential
conductance between terminal 1 and 3 when no current flows through terminal 2,
i.e., $G_1=\left.\frac{dI_1}{dV_1}\right|_{I_2=0}$; (ii) the differential 
conductance between terminal 2 and 3 when no current flows through terminal 1,
i.e., $G_2=\left.\frac{dI_2}{dV_2}\right|_{I_1=0}$. 
Generally speaking, these two quantities cannot be obtained through a simple 
analytic formula but can only calculated through a self-consistent numerical
scheme. We take $G_1$ as an example to explain this scheme:
(1) for certain $V_1$, $\mu_\eta$ and $I_\eta$
can be determined self-consistently using the conditions $I_2=0$ and 
$\sum_\eta I_\eta=0$; (2) for the bias slightly deviating from $V_1$, termed as 
$V_1'$, one obtains the corresponding current $I_1'$ in a similar way;
(3) the differential conductance is obtained from $(I_1'-I_1)/(V_1'-V_1)$.

When all arms of the $T$-shaped Majorana nanostructure are very long, the
transmissions between different leads (i.e., $P_{\eta\eta'}^{e\beta}$ for
$\eta\ne\eta'$) become negligible around zero energy due to the presence of
the superconducting gap and hence only the Andreev reflection contributes to the 
transport. In this case, Eq.~(\ref{current_2}) becomes simpler, 
\begin{equation}
  I_\eta =\frac{e}{h} \int_{-\mu_{\eta}}^{\mu_\eta} d\varepsilon \;
  P_{\eta\eta}^{eh}(\varepsilon).
\end{equation}
Further using $P_{11}^{eh}(\varepsilon)=P_{33}^{eh}(\varepsilon)$, which comes
from the left-right symmetry of this structure, 
one obtains the conductance $G_1$ 
\begin{equation}
  G_1={e^2}/{2h} \left[ P_{11}^{eh}\left({eV_1}/{2}\right)
    + P_{11}^{eh}\left(-{eV_1}/{2}\right) \right].
  \label{conduct_simple}
\end{equation}
Nevertheless, the conductance $G_2$ in this case still needs to be obtained
through a self-consistent scheme.


\subsection{Numerical results}
\begin{figure}[tbp]
  \begin{center}
    \includegraphics[width=7.cm]{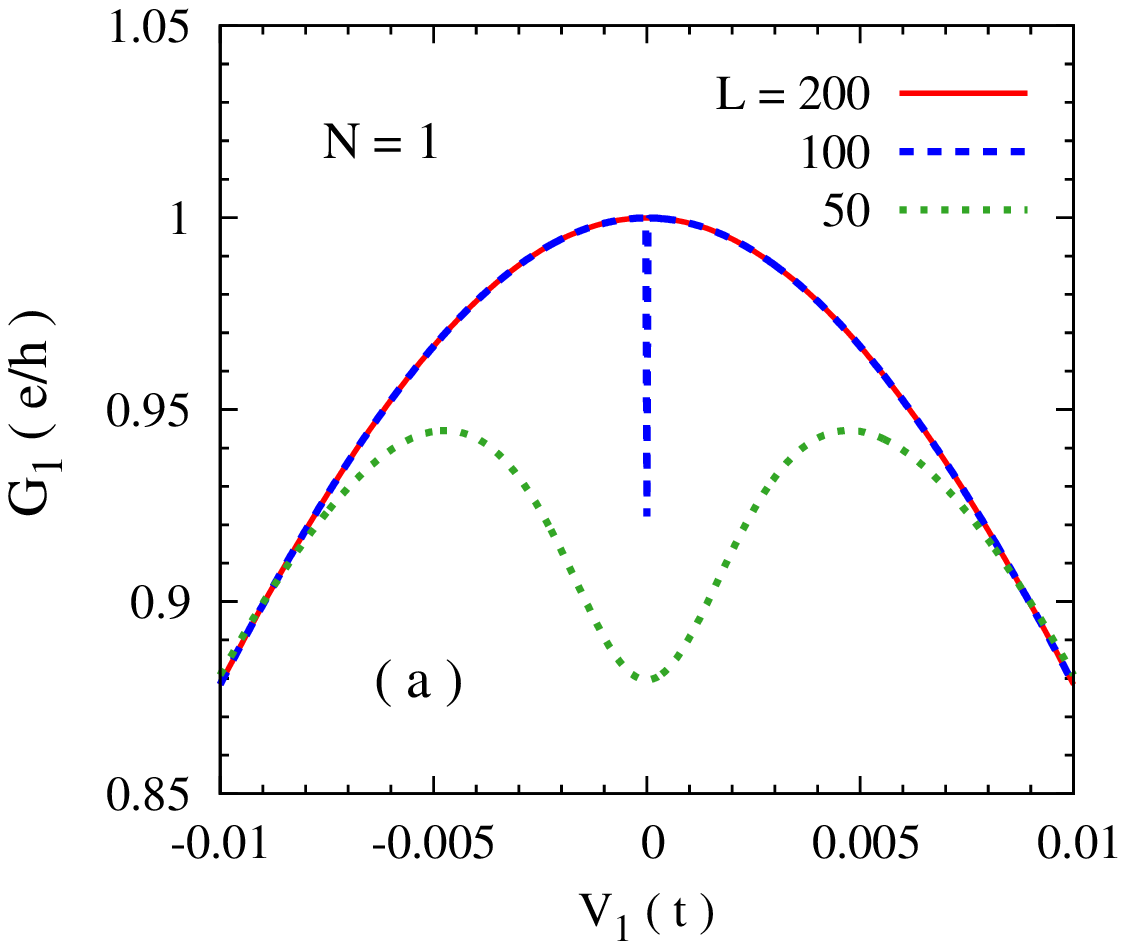}
    \includegraphics[width=7.cm]{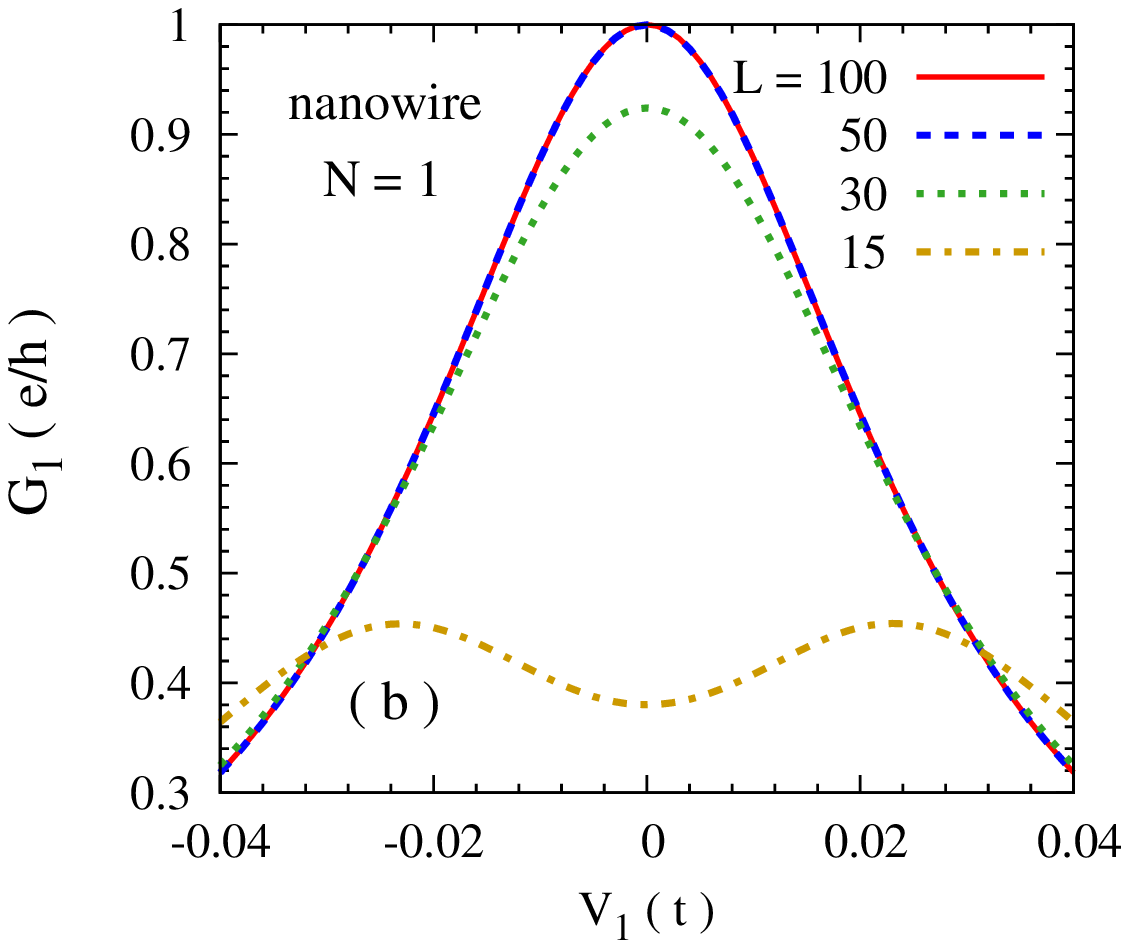}
    \includegraphics[width=7.cm]{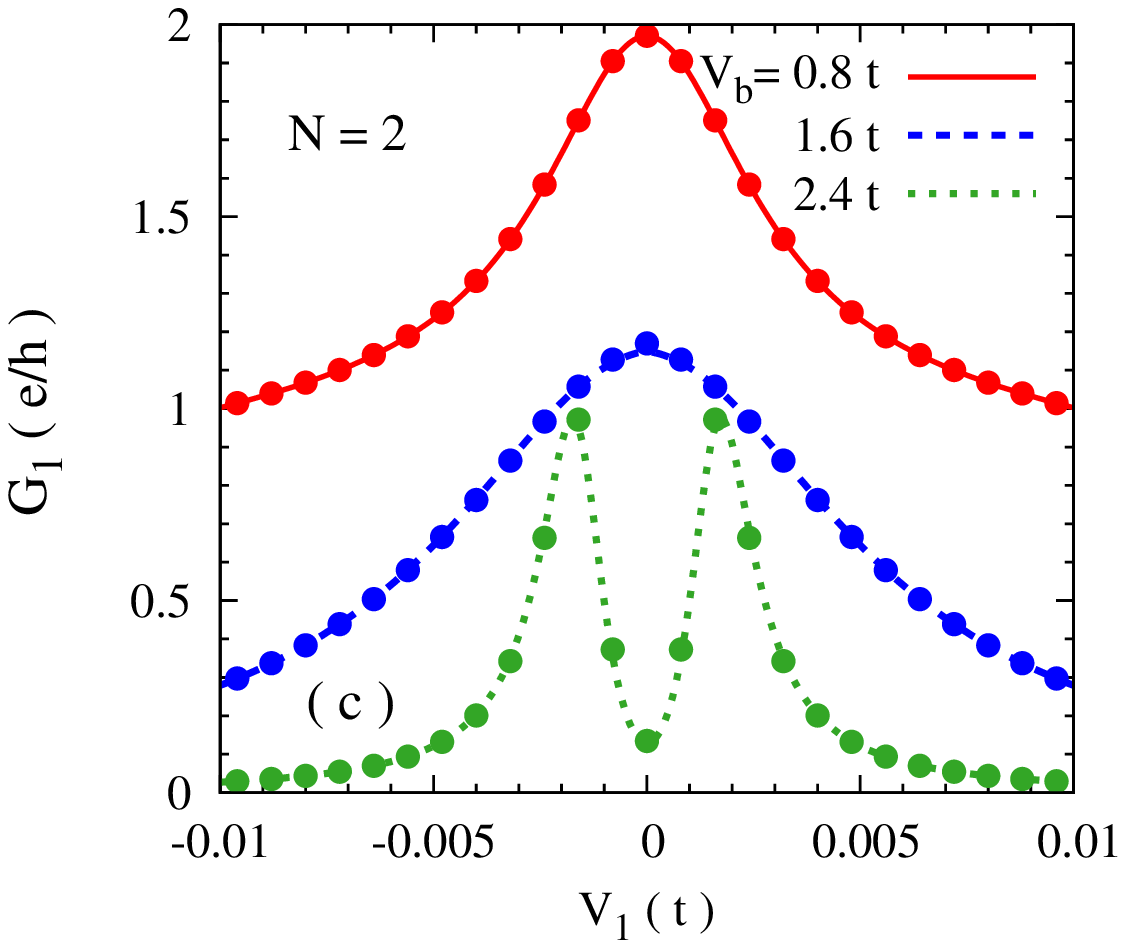}
  \end{center}
  \caption{(Color online) Conductance $G_1$ in $T$-shaped Majorana
    nanostructures (a) and nanowires (b) versus bias $V_1$ with different 
    lengths for the Zeeman splitting $V_Z=0.8 t$ (corresponding to $N=1$) and
    barrier height $V_b=0.8 t$. 
    (c) Conductance $G_1$ in $T$-shaped Majorana nanostructures (curves)
    and nanowires (dots) versus bias $V_1$ with different barrier heights for
    $V_Z=1.2 t$ (corresponding to $N=2$) and $L=100$. 
  }
  \label{fig_conduct} 
\end{figure}

In this subsection, we present the numerical results of the conductance through
the $T$-shaped Majorana structure in various parameter regimes. 
We choose $W=W_s=4$ and $\mu=0.4 t$ just as Fig.~\ref{fig_multi}.
We also set the barrier width $W_b=2$ throughout this subsection.
We first discuss the bias dependence of conductance for $V_{\rm Z}=0.8 t$, which
belongs to the regime with $N=1$, as shown in Fig.~\ref{fig_phase}. 
The conductance $G_1$ is plotted against bias with different arm lengths
for the barrier height $V_b=0.8 t$ in Fig.~\ref{fig_conduct}(a).
The behavior of $G_2$ is similar to this one and not shown here.
It is seen that the conductance exhibits a Lorentzian peak
at zero bias with the peak value being $e^2/h$ for $L=L_s=200$.
This is just the typical behavior of Majorana fermion-assisted 
transport,\cite{Lim_trans_NSN} indicating the arm has been long enough so that
the interaction between different Majorana modes becomes negligible. 
The behaviors with shorter arm are more interesting. 
One observes a  sharp valley at zero bias and double peaks at
finite bias for $L=L_s=100$ and $50$. 
Note that these behaviors are very distinct from those in  
nanowires, which are plotted in Fig.~\ref{fig_conduct}(b) with the
same parameters as the previous ones except $L_s=0$ (i.e., the side-arm is
removed). In that figure, one observes that the conductance shows the
double-peak structure only for extremely short length $L=15$. Obviously, the
double-peak behavior appears at much longer length in the $T$-shaped structure
compared with that in the nanowire.

Two reasons lead to the above distinct behaviors in these two
structures. The first one is straightforward: due to the presence of the
intersection Majorana mode, the distance of the adjoining Majorana modes $L$ in
the $T$-shaped structure is only about one half of that in the 
corresponding nanowire, whose total length is $2L+W_s$. 
This enhances the interaction between the adjoining Majorana modes and makes the 
split-peak structure appear at longer length. 
The second reason is more subtle: all the Majorana modes in the nanowire are
located at the ends and hence their self-energies from the leads are large,
whereas the intersection Majorana mode in the $T$-shaped structure has a
very small self-energy. 
The influence of this factor can be seen clearly in the limit where the
self-energy of the intersection Majorana mode is negligible compared with all
the other quantities. In this limit, the conductance can be described by
Eq.~(\ref{G_model}) in Appendix~\ref{current_approx} with
$\Gamma_{1}=\Gamma_{1L}$ and $\Gamma_{2}=\Gamma_{2L}=0$ (the Majorana modes at
left end and the intersection are numbered 1 and 2, respectively) 
\begin{equation}
  G_1(V_1)=\frac{e^2}{h} \frac{\Gamma_{1}^2e^2 V_1^2}
  {(e^2 V_1^2-4|\varepsilon_{12}|^2)^2+\Gamma_1^2e^2 V_1^2}.
  \label{G_1_valley2}
\end{equation}
From this formula, one finds that $G_1(V_1)$ takes its minimum
value $0$ at zero bias and reaches its maximum value $e^2/h$ at $eV_1=\pm
2|\varepsilon_{12}|$ with $|\varepsilon_{12}|$ representing the interaction
between the adjoining Majorana modes. This indicates that the conductance always
shows the double-peak structure in this limit. 
In fact, the self-energy of the intersection Majorana mode is not so
small in most cases and the behavior of the conductance in the $T$-shaped
structure is usually between the above limit and the Lorentzian-peak
behavior. Nevertheless, the small self-energy of the intersection mode still 
facilitates the formation of the split-peak behavior and makes it appear at
longer length. 


Then we turn to the bias dependence of conductance for $V_{\rm Z}= 1.2 t$,
which corresponds to $N=2$, i.e., there are two near-zero-energy modes at each
ends. As discussed in Sec.~II, the splitting of these modes is mainly from the
inter-subband SOC and hence insensitive to the arm length as long as the arm is
not too short. 
Thus, here we do not change the length as the previous case with $N=1$, instead,
we fix the length $L=L_s=100$ and change the barrier height to show the typical
transport behavior in this situation. 
We again only plot $G_1$ due to the similar behaviors between the conductances
$G_1$ and $G_2$. 
The results are plotted as curves in Fig.~\ref{fig_conduct}(b). 
It is seen that the conductance shows a peak at zero bias with the peak value
being close to $2e^2/h$ for low barrier height $V_b=0.8 t$, whereas exhibits
double peaks at finite bias when barrier height is large enough, e.g., 
$V_b=2.4 t$. The underlying physics is as follows.
For low barrier height, the self-energies of the near-zero-energy states are
larger than the splitting induced by the inter-subband SOC and hence all the
near-zero-energy modes just act as the same as the Majorana modes. 
In this case, the conductance can be described by Eq.~(\ref{G_model}) 
with $|\varepsilon_{12}|=0$, $\Gamma_{1}=\Gamma_{1L}$ and
$\Gamma_{2}=\Gamma_{2L}$,
\begin{equation}
  G_1(V_1)=\frac{e^2}{h} \left(\frac{\Gamma_1^2}{e^2 V_1^2+\Gamma_1^2}
  +\frac{\Gamma_2^2}{e^2 V_1^2+\Gamma_2^2}\right).
  \label{G_1_ideal}
\end{equation}
Evidently, $G_1$ in this case is just the summation of two Lorentzian functions
with height $e^2/h$.
For high barrier height, the self-energy of one of the low-energy modes at the 
end is much smaller than the splitting. Thus, the conductance can be
described by Eq.~(\ref{G_1_valley2}) and shows the split-peak behavior.
Moreover, we also plot the conductances through nanowires in the corresponding
cases as dots. It is seen that they almost coincide with the corresponding ones in
$T$-shaped structures. This is because in the case with $N=2$, there is no
low-energy states at the intersection and the properties of the low-energy
states at the three ends in $T$-shaped structures are similar to those in
nanowires.  
In addition, we also investigate the conductance for higher $N$ (not shown)
and find that the above phenomena in the case with $N=1$ ($2$) also 
appear in the case with $N$ being other odd (even) number.

\begin{figure}[tbp]
  \begin{center}
    \includegraphics[width=8.cm]{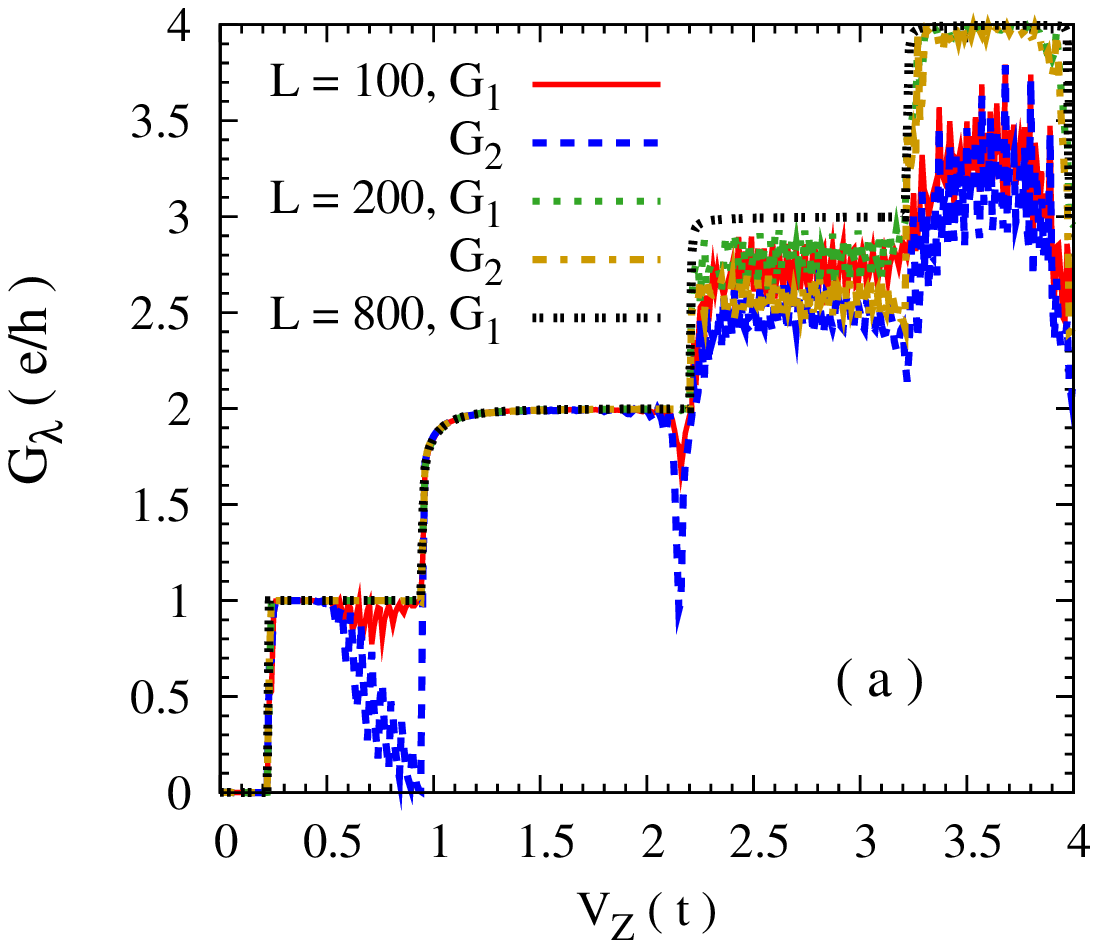}
    \includegraphics[width=8.cm]{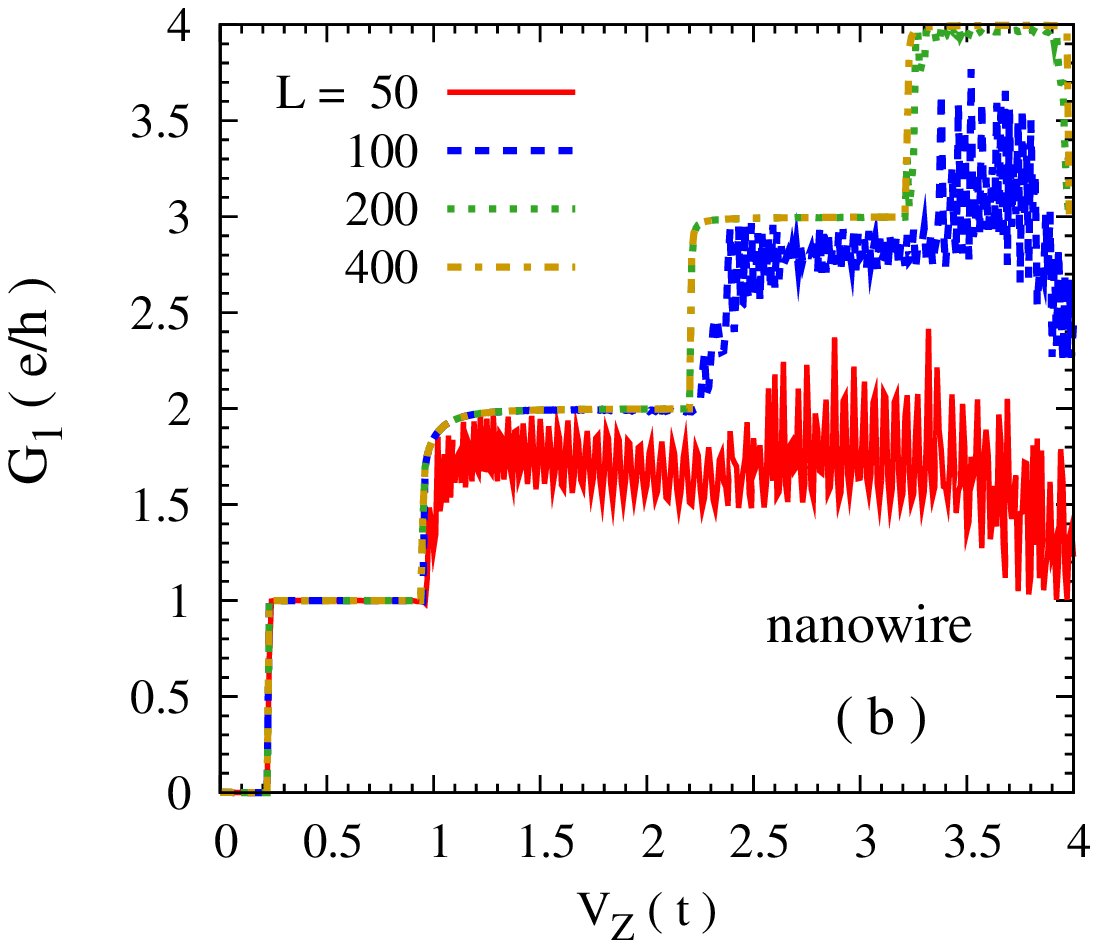}
  \end{center}
  \caption{\Black{(Color online) Linear conductances $G_1$ and $G_2$ in 
    $T$-shaped Majorana nanostructures (a) and $G_1$ in nanowires (b) versus
    Zeeman splitting with different lengths for the barrier height $V_b=0.8 t$. }
  }
  \label{fig_conduct2} 
\end{figure}

The unique transport properties in the $T$-shaped Majorana nanostructure 
can be seen more clearly in the magnetic-field dependence of the 
linear conductance (i.e., at zero bias).
Since the effect of the inter-subband SOC on the transport in the $T$-shaped
structure is similar to that in the nanowire addressed in the
literature,\cite{Lim_trans_NSN} here we focus on the case with $V_b=0.8 t$, 
where the splitting of the near-zero-energy states induced by the
inter-subband SOC is unimportant compared with their self-energies, as shown in 
Fig.~\ref{fig_conduct}(c). 
We plot $G_1(0)$ and $G_2(0)$ as function of magnetic 
field for different arm lengths in Fig.~\ref{fig_conduct2}(a). 
We first discuss the case in the long-arm limit, i.e., $L=800$.
As $G_1$ and $G_2$ coincide in this case, only $G_1$ is shown. 
It is seen that the linear conductance is very close
to the ideal value $Ne^2/h$ in all parameter regimes 
investigated in this work. This indicates that the splitting of the relevant 
low-energy modes is negligible compared with their self-energies. 

The behavior becomes more interesting for shorter arm length.
In the case with $L=200$, both $G_1$ and $G_2$ take the ideal value
$Ne^2/h$ in the  regimes for $N=1$, 2 and 4, however, deviate
much from their ideal value for $N=3$. 
This phenomenon can be understood as follows.
As said above, in the regime for odd $N$, the intersection Majorana mode
emerges, which enhances the discrepancy between the zero-bias conductance and
its ideal value. 
It is also known that, with the increasing magnetic field, the coherence length
of the low-energy modes tends to increase, and hence the splitting of these
states tends to increase.\cite{Sarma_smoking} 
Therefore, the pronounced deviation from the ideal value appears in the regime
for $N=3$, in which $N$ is odd and the corresponding magnetic field is high. 
It is also shown that the deviation in $G_2$ is larger than
$G_1$. This can be understood by considering the magnitude of the wave
function of the intersection Majorana mode in the side-arm is larger than that 
in the main-arm, as shown in its analytic solution with $W=W_s=1$
[Eqs.~(\ref{wave_center_1})-(\ref{wave_center_3})]. 
Note that the deviation from the ideal value in the regime for $N=3$ is
even larger than that for $N=4$, although the latter one appears at higher
magnetic field. 
Similar phenomenon is observed in the case with $L=50$. In that case, one finds
that the derivation for $N=1$ is larger than that for $N=2$. 
The above behaviors are very distinct from those in Majorana nanowires.
In that system, as shown in Fig.~\ref{fig_conduct2}(b), the deviation from the
ideal value always tends to increase with the increase of magnetic field due to
the decrease of the coherence length.


\section{Conclusion}
In conclusion, we have investigated the Majorana fermions in a $T$-shaped
semiconductor nanostructure with the Rashba SOC and proximity-induced
superconducting pairing in the presence of a magnetic field perpendicular to the 
plane of this structure. 
We first discuss the low-energy spectrum of this system.
We find that the properties of the low-energy modes (including the Majorana
and near-zero-energy modes) at the ends of the $T$-shaped structure are
similar to those in the Majorana nanowire. 
The number of the low-energy modes at each end $N$ is approximately equal to the number
of the subbands in which only the states with one kind of spin are occupied and
the number of the Majorana modes at each end is one (zero) for odd (even) $N$.
Moreover, very distinct from the nanowire, it is discovered that one Majorana mode
appears at the intersection of the $T$-shaped structure in the case with odd $N$
to ensure that the total number of the Majorana modes is even. However, there is
neither Majorana nor near-zero-energy mode at the intersection for even $N$.

We also investigate the transport properties through the above $T$-shaped
nanostructure with each end connected with a normal lead.
It is found that the deviation of the zero-bias conductance from its
ideal value in the long-arm limit $Ne^2/h$ is more pronounced in the regime for
odd $N$ compared to the one for even $N$. 
This is because the presence of the intersection Majorana mode reduces the
distance between the adjoining Majorana modes and the self-energy of this
intersection mode from the leads is very small. 
Moreover, the regime for odd $N$ can appear at
lower magnetic field than that for even one. Therefore, around the boundary
between these two regimes, the deviation from the ideal value tends to decrease
with increasing magnetic field. 
This behavior is also very distinct from that in the nanowire, where
the deviation from the ideal value always tends to increase with increasing
magnetic field due to the decrease of the coherence 
length of the low-energy modes.

\begin{acknowledgments}
  This work was supported by the National Natural Science Foundation of China 
  under Grant No.\ 11334014, the National Basic Research Program of China under
  Grant No.\ 2012CB922002 and the Strategic Priority Research Program of the
  Chinese Academy of Sciences under Grant No.\ XDB01000000. 
\end{acknowledgments}

\begin{appendix}
\section{Wave functions of Majorana modes in one-dimensional 
  $T$-shaped structure} 
\label{MF_spe}
In this appendix, we present the derivation of the wave functions of the
Majorana modes in the one-dimensional $T$-shaped structure, i.e., $W=W_s=1$.
It is known that, in the nontrivial topological regime, one
Majorana mode appears at each end of the one-dimensional nanowire. This
indicates that there is one solution $\Phi_0(x)$ satisfying the
BdG equation ${H}_{\rm BdG}(x',x)\Phi_0(x)=0$ and the boundary condition
$\Phi_0(0)=0$. Also from the particle-hole symmetry of the BdG Hamiltonian, 
any zero-energy solution can be written into the form
$\Phi_\eta(x)=(u_\eta(x), i\hat{\sigma}_y u_\eta(x))^T$.\cite{Sarma_smoking} 
Thus, $u_0(x)$ corresponds to the above solution $\Phi_0(x)$. 
After performing the translation and rotation, one obtains the normalized wave
function of the Majorana mode $u_\eta(x,y)$ at the end $\eta$ of the $T$-shaped  
nanostructure (see Fig.~\ref{fig_structure}),
\begin{eqnarray}
  u_1(x,y)&=& u_0(x+L) \delta_{y,0},\\
  u_2(x,y)&=&\frac{\sqrt{2}}{2}(\hat{I}+i\hat{\sigma}_z) u_0(-y+L) \delta_{x,0},\\
  u_3(x,y)&=&i\hat{\sigma}_z u_0(-x+L) \delta_{y,0}. 
\end{eqnarray}
Generally speaking, the wave function of the intersection Majorana mode
$u_4(x,y)$ cannot be constructed in this way. However, for $W=W_s=1$, the exact
numerical calculation gives $u_4(x=0,y=0)=0$ within the computational accuracy.
Thus one obtains the form of the intersection Majorana mode,
\begin{eqnarray}
  u_4(x>0,y=0)&=&\frac{A}{2} u_0(x),
  \label{wave_center_1}\\
  u_4(x<0,y=0)&=& \frac{iB}{2} \hat{\sigma}_z u_0(-x),
  \label{wave_center_2}\\
  u_4(x=0,y>0)&=&\frac{C}{2}(\hat{I}-i\hat{\sigma}_z)u_0(y).
  \label{wave_center_3}
\end{eqnarray}
It can be verified that the above solution $u_4(x,y)$ indeed satisfies the
BdG equation when $A=-B=-C=1$.

\section{Green's functions in Nambu spinor basis}
\label{Green_Func}
Here we briefly discuss the Green's functions in the Nambu spinor
basis.\cite{Sarma_single,Sarma_multi,Potter_multi,Gibertini_multi} 
We first define the contour-ordered Green's functions 
in a isolated superconducting nanostructure in this basis as
\begin{equation}
  g_{i\sigma j\sigma'}^{c,\alpha\beta}(t,t') = -i \langle T_c\,
  \tilde{c}_{\alpha i\sigma}(t) \tilde{c}_{\beta j\sigma'}^\dagger(t') \rangle  
  \label{g_c}
\end{equation}
with $\tilde{c}_{e{i\sigma}} \equiv c_{{i\sigma}}$ and 
$\tilde{c}_{h{i\sigma}} \equiv \sigma c_{{i-\sigma}}^\dagger$.
After connecting this superconducting structure with normal leads, the
contour-ordered Green's functions can be obtained through the Dyson equation 
\begin{equation}
  \hat{G}^c(t,t')=\hat{g}^c(t,t') + \int_c dt_1 dt_2 \hat{g}^c(t,t_1)
  \hat{\Sigma}^c(t_1,t_2) \hat{G}^c(t_2,t').
  \label{Dyson_c}
\end{equation}
Here symbols with hat (\,$\hat{}$\,) represent the corresponding quantities in
the lattice and Nambu spinor space; $\hat{\Sigma}^c(t_1,t_2)$ denotes the total
self-energy from all leads 
\begin{eqnarray}
  \Sigma^{c,\alpha\beta}_{i_1\sigma_1,i_2\sigma_2}(t_1,t_2)
  &=&\sum_{\eta j_1\sigma_1' j_2 \sigma_2'}
  T^{\eta,\alpha}_{i_1\sigma_1, j_1\sigma_1'} 
  T^{\eta,\alpha}_{j_2 \sigma_2', i_2\sigma_2} 
  \nonumber \\
  && {} \times F^{c,\eta,\alpha\alpha}_{j_1\sigma_1', j_2 \sigma_2'}(t_1,t_2) 
  \delta_{\alpha\beta},
  \label{Sigma_def}
\end{eqnarray}
in which
\begin{eqnarray}
  F^{c,\eta,\alpha\beta}_{j\sigma, j \sigma'}(t,t')
  &=& -i \langle T_c\, {d}_{\eta,\alpha,i\sigma}(t) 
  {d}_{\eta,\beta,j\sigma'}^\dagger(t') \rangle, \\
  T^{\eta,\alpha}_{i\sigma, j\sigma'}&=& \left\{ \begin{array}{cc} 
      T^{\eta}_{i\sigma, j\sigma'} & \alpha=e \\ 
      \sigma \sigma' T^{\eta\,\ast}_{i-\sigma, j-\sigma'}
      & \alpha=h \end{array} \right. .
\end{eqnarray}

Similar to Eq.~(\ref{g_c}), one can define the retarded, advanced,
lesser and greater Green's functions in the
isolated superconducting nanostructure $\hat{g}^{r,a,<,>}(t,t')$.
Further performing the Fourier transformation, one obtains
$\hat{g}^{r,a,<,>}(\varepsilon)$. 
It can be demonstrated that these Green's functions satisfy
\begin{eqnarray}
  && (\varepsilon- \hat{H}_{\rm BdG} + i0^+) \hat{g}^r(\varepsilon)
  = 1,
  \label{g_r_eq}\\
  && (\varepsilon- \hat{H}_{\rm BdG}) \hat{g}^<(\varepsilon)= 0.
  \label{g_l_eq}
\end{eqnarray}
Since the investigated system is finite, the infinitesimal in
Eq.~(\ref{g_r_eq}) can be neglected.\cite{Book_nonequilibrium,Zhou_infinite}
Thus, 
\begin{eqnarray}
  && [\hat{g}^r(\varepsilon)]^{-1}-[\hat{g}^a(\varepsilon)]^{-1}=0,
  \label{g_r_eq_2}\\
  && [\hat{g}^r(\varepsilon)]^{-1} \hat{g}^<(\varepsilon)=0.
  \label{g_l_eq_2}
\end{eqnarray}
Performing the Langreth rules\cite{Haug_1998} and the Fourier transformation on 
Eq.~(\ref{Dyson_c}) and further exploiting 
Eqs.~(\ref{g_r_eq_2}) and (\ref{g_l_eq_2}), one obtains
\begin{eqnarray}
  && \hspace{0.75cm} \hat{G}^<(\varepsilon) =
  \hat{G}^r(\varepsilon) \hat{\Sigma}^<(\varepsilon)
  \hat{G}^a(\varepsilon), 
  \label{g_r_relation} \\
  && \hspace{-0.5cm} 
  \hat{G}^r(\varepsilon)-\hat{G}^a(\varepsilon)
  = \hat{G}^r(\varepsilon) [\hat{\Sigma}^r(\varepsilon)
    -\hat{\Sigma}^a(\varepsilon)] \hat{G}^a(\varepsilon).
  \label{g_l_relation}
\end{eqnarray}
Note that the above relations are in the same form as those well-known relations 
for the Green's functions in the normal
conductor.\cite{Haug_1998,Book_nonequilibrium} 
This indicates that the formula of current through the superconducting 
mesoscopic nanostructure can be derived following
the similar way to the current through the normal nanostructure.\cite{Haug_1998}

\section{Approximate formula of conductance induced by two interacting 
  Majorana modes} 
\label{current_approx}
When only two Majorana modes contribute to the low-energy transport,
all Green's functions and self-energies can be reduced into the small space
formed by these two modes. Then one obtains
\begin{eqnarray}
  &&\hat{G}^{r}(\varepsilon)=\left[\varepsilon-\begin{pmatrix} 0 & 
      \varepsilon_{12}\\ \varepsilon_{12}^\ast & 0\end{pmatrix}
    -\frac{i}{2} \begin{pmatrix} \Gamma_1 & 0\\ 
      0 & \Gamma_2 \end{pmatrix} \right]^{-1}, 
  \label{G_r_spe} \\
  &&\hat{\Gamma}^{e}_{1}(\varepsilon) =
  \hat{\Gamma}^{h}_{1}(\varepsilon) = \frac{1}{2} \begin{pmatrix} 
    \Gamma_{1L} & 0\\ 0 & \Gamma_{2L} \end{pmatrix}.
  \label{Gamma_spe}
\end{eqnarray}
Here $\varepsilon_{12}$ represents the coupling between the two Majorana modes;
$\Gamma_i$ and $\Gamma_{iL}$ stand for the total self-energy and 
the one from the left lead of the $i$-th Majorana mode, respectively.
Substituting Eqs.~(\ref{G_r_spe}) and (\ref{Gamma_spe}) into Eqs.~(\ref{P_def})
and (\ref{conduct_simple}), one obtains 
\begin{eqnarray}
  \nonumber
  G_1(V_1) &=&\frac{e^2}{h}[(\Gamma_{1L}^2+\Gamma_{2L}^2) e^2 V_1^2
  +\Gamma_1^2\Gamma_{2L}^2+\Gamma_{1L}^2\Gamma_{2}^2 \\
  \nonumber
  &&{}+8\Gamma_{1L}\Gamma_{2L}|\varepsilon_{12}|^2]
  [(e^2 V_1^2-4|\varepsilon_{12}|^2-\Gamma_1\Gamma_2)^2 \\
  &&{}+(\Gamma_1+\Gamma_2)^2e^2 V_1^2]^{-1}.
  \label{G_model}
\end{eqnarray}
\end{appendix}


\begin{thebibliography}{0}
\bibitem{upper_limit} Note that the magnitude of the Zeeman splitting 
  of the nontrivial topological regime has an upper limit in the tight-binding
  model,\cite{Gibertini_multi} which is distinct from the continuum model.
\end{thebibliography}
\end{document}